\documentclass[final,numbers]{aipproc}
\usepackage{amssymb}

\layoutstyle{8x11double}

\begin{document}

\title{The ``fireshell'' model and the ``canonical'' GRB scenario.}

\classification{98.70.Rz}
\keywords      {Gamma-Ray: Bursts}

\author{Carlo Luciano Bianco}{address={ICRANet, Piazzale della Repubblica 10, 65122 Pescara, Italy.}, altaddress={Dipartimento di Fisica, Universit\`a di Roma ``La Sapienza'', Piazzale Aldo Moro 5, 00185 Roma, Italy.}}

\author{Maria Grazia Bernardini}{address={ICRANet, Piazzale della Repubblica 10, 65122 Pescara, Italy.}, altaddress={Dipartimento di Fisica, Universit\`a di Roma ``La Sapienza'', Piazzale Aldo Moro 5, 00185 Roma, Italy.}}

\author{Letizia Caito}{address={ICRANet, Piazzale della Repubblica 10, 65122 Pescara, Italy.}, altaddress={Dipartimento di Fisica, Universit\`a di Roma ``La Sapienza'', Piazzale Aldo Moro 5, 00185 Roma, Italy.}}

\author{Maria Giovanna Dainotti}{address={ICRANet, Piazzale della Repubblica 10, 65122 Pescara, Italy.}, altaddress={Dipartimento di Fisica, Universit\`a di Roma ``La Sapienza'', Piazzale Aldo Moro 5, 00185 Roma, Italy.}}

\author{Roberto Guida}{address={ICRANet, Piazzale della Repubblica 10, 65122 Pescara, Italy.}, altaddress={Dipartimento di Fisica, Universit\`a di Roma ``La Sapienza'', Piazzale Aldo Moro 5, 00185 Roma, Italy.}}

\author{Remo Ruffini}{address={ICRANet, Piazzale della Repubblica 10, 65122 Pescara, Italy.}, altaddress={Dipartimento di Fisica, Universit\`a di Roma ``La Sapienza'', Piazzale Aldo Moro 5, 00185 Roma, Italy.}}

\begin{abstract}
In the ``fireshell'' model we define a ``canonical GRB'' light curve with two sharply different components: the Proper-GRB (P-GRB), emitted when the optically thick fireshell of electron-positron plasma originating the phenomenon reaches transparency, and the afterglow, emitted due to the collision between the remaining optically thin fireshell and the CircumBurst Medium (CBM). We outline our ``canonical GRB'' scenario, originating from the gravitational collapse to a black hole, with a special emphasis on the discrimination between ``genuine'' and ``fake'' short GRBs.
\end{abstract}

\maketitle

\section{Introduction}

The observations of GRB 060614 (\citet{ge06,ma07}) challenged the standard GRB classification scheme (\citet{k92,da92}) in which the gamma events are branched into two classes: ``short'' GRBs (events which last less than $\sim 2$s) and ``long'' GRBs (events which last more than $\sim 2$s). GRB 060614, indeed, ``reveals a first short, hard-spectrum episode of emission (lasting $5$ s) followed by an extended and somewhat softer episode (lasting $\sim 100$ s)'': a ``two-component emission structure'' (\citet{ge06}). Moreover, stringent upper limits on the luminosity of the Supernova possibly associated with GRB 060614 have been established (\citet{da06,gya06}). \citet{ge06} concluded that ``it is difficult to determine unambiguously which category GRB 060614 falls into'' and that, then, GRB 060614, due to its ``hybrid'' observational properties, ``opens the door on a new GRB classification scheme that straddles both long and short bursts'' (\citet{ge06}).

These observations motivated \citet{nb06} to reanalyze the BATSE catalog identifying a new GRB class with ``an occasional softer extended emission lasting tenths of seconds after an initial spikelike emission'' (\citet{nb06}). In some cases, ``the strength of the extended emission converts an otherwise short burst into one with a duration that can be tens of seconds, making
it appear to be a long burst'' (\citet{nb06}). Hence, \citet{nb06} suggested that the standard ``long-short'' GRB classification scheme ``is at best misleading'' (\citet{nb06}).

In the following, we are going to outline our ``canonical GRB'' scenario (\citet{rlet1,rlet2,XIIBSGC,970228}), where all GRBs are generated by the same ``engine'': the gravitational collapse to a black hole. We will show that such ``hybrid'' sources are indeed explainable in terms of a peculiarly small average value of the CircumBurst Medium (CBM) density, compatible with a galactic halo environment (see \citet{970228,970228_IC4}).

\section{The ``fireshell'' model}

We assume that all GRBs, including the ``short'' ones, originate from the gravitational collapse to a black hole (\citet{rlet2,XIIBSGC}). The $e^\pm$ plasma created in the process of the black hole formation expands as an optically thick and spherically symmetric ``fireshell'' with a constant width in the laboratory frame, i.e. the frame in which the black hole is at rest. We have only two free parameters characterizing the source, namely:
\begin{itemize}
\item $E_{e^\pm}^{tot}$: the total energy of the $e^\pm$ plasma,
\item $B\equiv \displaystyle\frac{M_Bc^2}{E_{e^\pm}^{tot}}\,$: the $e^\pm$ plasma baryon loading,
\end{itemize}
where $M_B$ is the total baryons' mass (\citet{rswx00}). These two parameters fully determine the optically thick acceleration phase of the fireshell, which lasts until the transparency condition is reached and the Proper-GRB (P-GRB) is emitted (\citet{rlet2}).

The afterglow emission then starts due to the collision between the remaining optically thin fireshell and the CBM (\citet{rlet2,EQTS_ApJL,EQTS_ApJL2,PowerLaws,XIIBSGC}. It clearly depends on the parameters describing the effective CBM distribution:
\begin{itemize}
\item $n_{cbm}$: its density,
\item ${\cal R}\equiv \displaystyle\frac{A_{eff}}{A_{vis}}$: its filamentary structure,
\end{itemize}
where $A_{eff}$ is the effective emitting area of the fireshell and $A_{vis}$ is its total visible area (\citet{rlet02,spectr1,fil,060218}).

\section{The ``canonical'' GRB scenario}

\begin{figure}
\includegraphics[width=\hsize,clip]{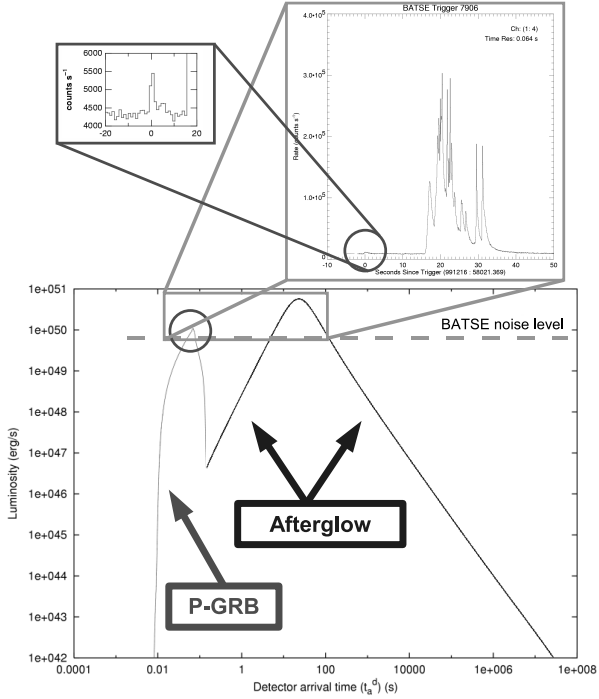}
\caption{The ``canonical GRB'' light curve theoretically computed for GRB 991216. The prompt emission observed by BATSE is identified with the peak of the afterglow, while the small precursor is identified with the P-GRB. For this source we have $B\simeq 3.0\times 10^{-3}$ and $\langle n_{cbm} \rangle \sim 1.0$ particles/cm$^3$. Details in \citet{rlet2,rlet02,rubr,rubr2}.}
\label{991216_fig}
\end{figure}

Unlike treatments in the current literature (see e.g. \citet{p04,m06} and references therein), we define a ``canonical GRB'' light curve with two sharply different components (see Fig. \ref{991216_fig} and \citet{rlet2,XIIBSGC,970228}):
\begin{enumerate}
\item \textbf{The P-GRB:} it has the imprint of the black hole formation, an harder spectrum and no spectral lag (\citet{brx01,rfvx05}).
\item \textbf{The afterglow:} it presents a clear hard-to-soft behavior (\citet{031203,spectr1,050315}); the peak of the afterglow contributes to what is usually called the ``prompt emission'' (see e.g.\citet{rlet2,050315,060218}).
\end{enumerate}

The ratio between the total time-integrated luminosity of the P-GRB (namely, its total energy) and the corresponding one of the afterglow is the crucial quantity for the identification of GRBs' nature. Such a ratio, as well as the temporal separation between the corresponding peaks, is a function of the $B$ parameter (\citet{rlet2}).

\begin{figure}
\includegraphics[width=\hsize,clip]{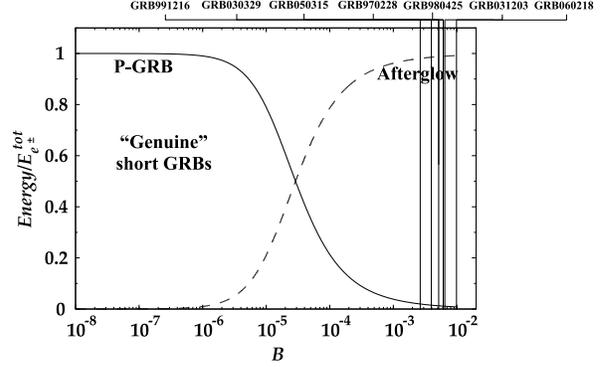}
\caption{The energy radiated in the P-GRB (solid line) and in the afterglow (dashed line), in units of the total energy of the plasma ($E_{e^\pm}^{tot}$), are plotted as functions of the $B$ parameter. Also represented are the values of the $B$ parameter computed for GRB 991216, GRB 030329, GRB 980425, GRB 970228, GRB 050315, GRB 031203, GRB 060218. Remarkably, they are consistently smaller than, or equal to in the special case of GRB 060218, the absolute upper limit $B \lesssim 10^{-2}$ established in \citet{rswx00}. The ``genuine'' short GRBs have a P-GRB predominant over the afterglow: they occur for $B \lesssim 10^{-5}$ (\citet{rlet2,970228}).}
\label{figX}
\end{figure}

When the P-GRB is the leading contribution to the emission and the afterglow is negligible we have a ``genuine'' short GRB (\citet{rlet2}). This is the case where $B \lesssim 10^{-5}$ (see Fig.~\ref{figX}): in the limit $B \to 0$ the afterglow vanishes (see Fig.~\ref{figX}).

\begin{figure}
\includegraphics[width=\hsize,clip]{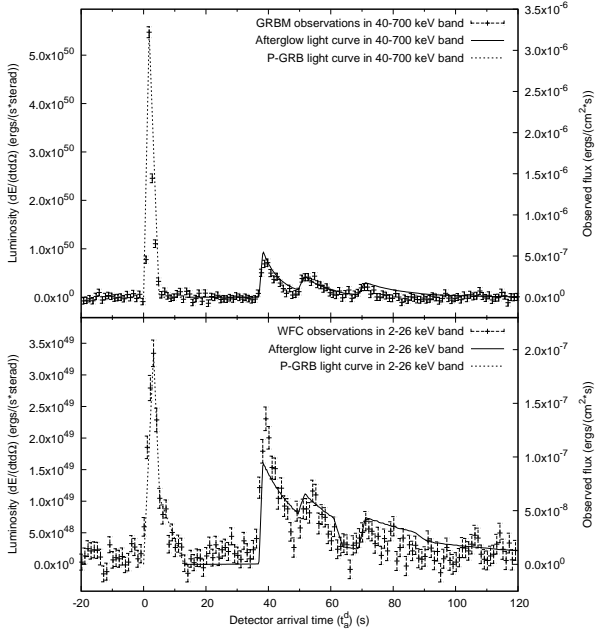}
\caption{The ``canonical GRB'' light curve theoretically computed for the prompt emission of GRB 970228. \emph{Beppo}SAX GRBM ($40$--$700$ keV, above) and WFC ($2$--$26$ keV, below) light curves (data points) are compared with the afterglow peak theoretical ones (solid lines). The onset of the afterglow coincides with the end of the P-GRB (represented qualitatively by the dotted lines). For this source we have $B\simeq 5.0\times 10^{-3}$ and $\langle n_{cbm} \rangle \sim 10^{-3}$ particles/cm$^3$. Details in \citet{970228,970228_IC4}.}
\label{970228_fit_prompt}
\end{figure}

In the other GRBs, with $10^{-4} \lesssim B \lesssim 10^{-2}$, the afterglow contribution is generally predominant (see Fig.~\ref{figX}; for the existence of the upper limit $B \lesssim 10^{-2}$ see \citet{rswx00} and \citet{060218}). Still, this case presents two distinct possibilities:
\begin{itemize}
\item The afterglow peak luminosity is \textbf{larger} than the P-GRB one. A clear example of this situation is GRB 991216, represented in Fig. \ref{991216_fig}.
\item The afterglow peak luminosity is \textbf{smaller} than the P-GRB one. A clear example of this situation is GRB 970228, represented in Fig. \ref{970228_fit_prompt}.
\end{itemize}

\begin{figure}
\includegraphics[width=\hsize,clip]{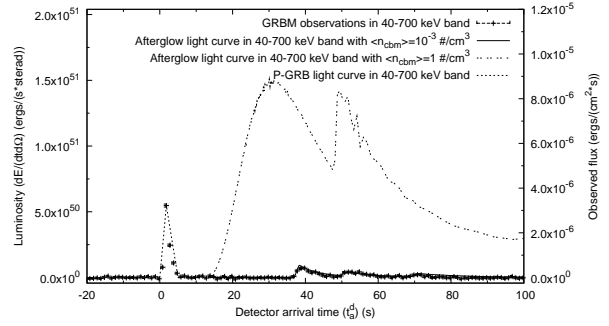}
\caption{The theoretical fit of the \emph{Beppo}SAX GRBM observations (solid line, see Fig. \ref{970228_fit_prompt}) is compared with the afterglow light curve in the $40$--$700$ keV energy band obtained rescaling the CBM density to $\langle n_{cbm} \rangle = 1$ particle/cm$^3$ keeping constant its shape and the values of the fundamental parameters of the theory $E_{e^\pm}^{tot}$ and $B$ (double dotted line). The P-GRB duration and luminosity (dotted line), depending only on $E_{e^\pm}^{tot}$ and $B$, are not affected by this process of rescaling the CBM density. Details in \citet{970228}.}
\label{picco_n=1}
\end{figure}

The simultaneous occurrence of an afterglow with total time-integrated luminosity larger than the P-GRB one, but with a smaller peak luminosity, is indeed explainable in terms of a peculiarly small average value of the CBM density, compatible with a galactic halo environment, and not due to the intrinsic nature of the source (see Fig. \ref{970228_fit_prompt} and \citet{970228,970228_IC4}). Such a small average CBM density deflates the afterglow peak luminosity. Of course, such a deflated afterglow lasts much longer, since the total time-integrated luminosity in the afterglow is fixed by the value of the $B$ parameter (see above and Fig. \ref{picco_n=1}). In this sense, GRBs belonging to this class are only ``fake'' short GRBs. This is GRB class identified by \citet{nb06}, which also GRB 060614 belongs to, and which has GRB 970228 as a prototype (\citet{970228,970228_IC4,060614}).

\section{Conclusions}

\begin{figure}
\includegraphics[width=\hsize,clip]{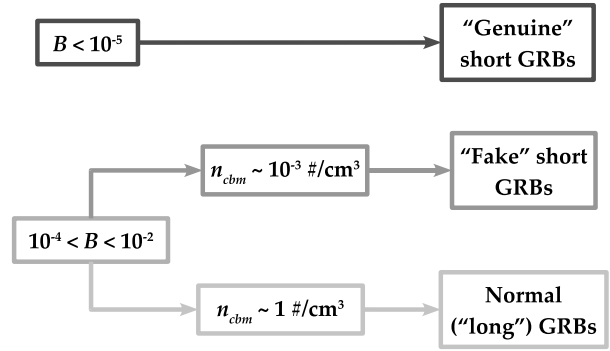}
\caption{A sketch summarizing the different possibilities predicted by the ``canonical GRB'' scenario depending on the fireshell baryon loading $B$ and the average CBM density $\langle n_{cbm} \rangle$.}
\label{canonical}
\end{figure}

We have presented our ``canonical GRB'' scenario, especially pointing out the need to distinguish between ``genuine'' and ``fake'' short GRBs:
\begin{itemize}
\item The \textbf{``genuine'' short GRBs} inherit their features from an intrinsic property of their sources. The very small fireshell baryon loading, in fact, implies that the afterglow time-integrated luminosity is negligible with respect to the P-GRB one.
\item The \textbf{``fake'' short GRBs}, instead, inherit their features from the environment. The very small CBM density, in fact, implies that the afterglow peak luminosity is lower than the P-GRB one, even if the afterglow total time-integrated luminosity is higher. This deflated afterglow peak can be observed as a ``soft bump'' following the P-GRB spike, as in GRB 970228 (\citet{970228,970228_IC4}), GRB 060614 (\citet{060614}), and the sources analyzed by \citet{nb06}.
\end{itemize}
A sketch of the different possibilities depending on the fireshell baryon loading $B$ and the average CBM density $\langle n_{cbm} \rangle$ is given in Fig. \ref{canonical}.

Before concluding, we turn to the Amati relation (\citet{aa02,amati06}) between the isotropic equivalent energy emitted in the prompt emission $E_{iso}$ and the peak energy of the corresponding time-integrated spectrum $E_p$. It clearly follows from our treatment (\citet{031203,spectr1,050315}) that both the hard-to-soft behavior and the Amati relation occurs uniquely in the afterglow phase which, in our model, encompass as well the prompt emission. The observations that the initial spikelike emission in the above mentioned ``fake'' short GRBs, which we identify with the P-GRBs, as well as all ``genuine'' short GRBs do not fulfill the Amati relation (see \citet{amati06}) is indeed a confirmation of our theoretical model (see also \citet{970228_IC4}). We look forward to verifications in additional sources.


\begin{thebibliography}{}

\bibitem[Amati(2006)]{amati06}
Amati, L. 2006, MNRAS, 372, 233.

\bibitem[Amati et al.(2002)]{aa02}
Amati, L., Frontera, F., Tavani, M., et al. 2002, A\&A, 390, 81.

\bibitem[Bernardini et al.(2007)]{970228}
Bernardini, M.G., Bianco, C.L., Caito, L., et al. 2007, A\&A, 474, L13.

\bibitem[Bernardini et al.(2007)]{970228_IC4}
Bernardini, M.G., Bianco, C.L., Caito, L., et al. 2007, AIP Conf. Proc., in this same volume.

\bibitem[Bernardini et al.(2005)]{031203}
Bernardini, M.G., Bianco, C.L., Chardonnet, P., et al. 2005, ApJ, 634, L29.

\bibitem[Bianco \& Ruffini(2004)]{EQTS_ApJL}
Bianco, C.L., Ruffini, R. 2004, ApJ, 605, L1.

\bibitem[Bianco \& Ruffini(2005a)]{EQTS_ApJL2}
Bianco, C.L., Ruffini, R. 2005, ApJ, 620, L23.

\bibitem[Bianco \& Ruffini(2005b)]{PowerLaws}
Bianco, C.L., Ruffini, R. 2005, ApJ, 633, L13.

\bibitem[Bianco et al.(2001)]{brx01}
Bianco, C.L., Ruffini, R., Xue, S.S. 2001, A\&A, 368, 377.

\bibitem[Caito et al.(2007)]{060614}
Caito, L., Bernardini, M.G., Bianco, C.L., et al. 2007, AIP Conf. Proc., in this same volume.

\bibitem[Dainotti et al.(2007)]{060218}
Dainotti, M.G., Bernardini, M.G., Bianco, C.L., et al. 2007, A\&A, 471, L29.

\bibitem[Della Valle et al.(2006)]{da06}
Della Valle, M., Chincarini, G., Panagia, N., et al. 2006, Nature, 444, 1050.

\bibitem[Dezalay et al.(1992)]{da92}
Dezalay, J.P., Barat, C., Talon, R., et al. 1992, AIP Conf. Proc. 265, 304.

\bibitem[Gal-Yam et al.(2006)]{gya06}
Gal-Yam, A., Fox, D.B., Price, P.A., et al. 2006, Nature, 444, 1053.

\bibitem[Gehrels et al.(2006)]{ge06}
Gehrels, N., Norris, J.P., Mangano, V., et al. 2006, Nature, 444, 1044.

\bibitem[Klebesadel(1992)]{k92}
Klebesadel, R.W. 1992, in ``Gamma-ray bursts'', CUP, p. 161.

\bibitem[Mangano et al.(2007)]{ma07}
Mangano, V., Holland, S.T., Malesani, D., et al. 2007, A\&A, 470, 105.

\bibitem[M\'esz\'aros(2006)]{m06}
M\'esz\'aros, P. 2006, Rep.Prog.Phys., 69, 2259.

\bibitem[Norris \& Bonnell(2006)]{nb06}
Norris, J.P, Bonnell, J.T., 2006, ApJ, 643, 266.

\bibitem[Piran(2004)]{p04}
Piran, T. 2004, Rev. Mod. Phys., 76, 1143.

\bibitem[Ruffini et al.(2005)]{rubr2}
Ruffini, R., Bernardini, M.G., Bianco, C.L., et al. 2005, AIP Conf. Proc. 782, 42.

\bibitem[Ruffini et al.(2006)]{050315}
Ruffini, R., Bernardini, M.G., Bianco, C.L., et al. 2006, ApJ, 645, L109.

\bibitem[Ruffini et al.(2007)]{XIIBSGC}
Ruffini, R., Bernardini, M.G., Bianco, C.L., et al. 2007, AIP Conf. Proc. 910, 55.

\bibitem[Ruffini et al.(2001)]{rlet1}
Ruffini, R., Bianco, C.L., Chardonnet, P., et al. 2001, ApJ, 555, L107.

\bibitem[Ruffini et al.(2001)]{rlet2}
Ruffini, R., Bianco, C.L., Chardonnet, P., et al. 2001, ApJ, 555, L113.

\bibitem[Ruffini et al.(2002)]{rlet02}
Ruffini, R., Bianco, C.L., Chardonnet, P., et al. 2002, ApJ, 581, L19.

\bibitem[Ruffini et al.(2003)]{rubr}
Ruffini, R., Bianco, C.L., Chardonnet, P., et al. 2003, AIP Conf. Proc. 668, 16.

\bibitem[Ruffini et al.(2004)]{spectr1}
Ruffini, R., Bianco, C.L., Chardonnet, P., et al. 2004, IJMPD, 13, 843.

\bibitem[Ruffini et al.(2005)]{fil}
Ruffini, R., Bianco, C.L., Chardonnet, P., et al. 2005, IJMPD, 14, 97.

\bibitem[Ruffini et al.(2005)]{rfvx05}
Ruffini, R., Fraschetti, F., Vitagliano, L., Xue, S.S., 2005, IJMPD, 14, 131.

\bibitem[Ruffini et al.(2000)]{rswx00} 
Ruffini, R., Salmonson, J.D., Wilson, J.R., Xue, S.S. 2000, A\&A, 359, 855.

\end{thebibliography}
\end{document}